\documentclass[pre,twocolumn,noshowpacs,amsmath,amssymb]{revtex4-2}
\usepackage{graphicx}
\usepackage{amsfonts}
\usepackage{dcolumn}
\usepackage{bm}
\usepackage{color}
\usepackage{hyperref}

\begin{document}
	
	\newcommand{\gin}[1]{{\bf\color{blue}#1}}
	\def\bc{\begin{center}}
		\def\ec{\end{center}}
	\def\bea{\begin{eqnarray}}
	\def\eea{\end{eqnarray}}
	\newcommand{\avg}[1]{\langle{#1}\rangle}
	\newcommand{\Avg}[1]{\left\langle{#1}\right\rangle}

\title{Epidemic extinction in a simplicial susceptible-infected-susceptible model}
\author{Yingshan Guo$^1$}
\author{Chuansheng Shen$^1$}\email{csshen@mail.ustc.edu.cn}
\author{Hanshuang Chen$^2$}\email{chenhshf@ahu.edu.cn}
\affiliation{$^1$School of Mathematics and Physics, Anqing Normal University, Anqing 246011, China \\ $^2$School of Physics and Optoelectronic Engineering, Anhui University, Hefei 230601, China }

\begin{abstract}
We study the extinction of epidemics in a simplicial susceptible-infected-susceptible model, where each susceptible individual becomes infected either by two-body interactions ($S+I \to 2I$) with a rate $\beta$ or by three-body interactions ($S+2I \to 3I$) with a rate $\beta (1+\delta)$, and each infected individual spontaneously recovers ($I \to S$) with a rate $\mu$. We focus on the case $\delta>0$ that embodies a synergistic reinforcement effect in the group interactions. By using the theory of large fluctuations to solve approximately for the master equation, we reveal two different scenarios for optimal path to extinction, and derive the associated action $\mathcal{S}$ for $\beta_b<\beta<\beta_c$ and for $\beta>\beta_c$, where $\beta_b=4 (1+\delta)/(2+\delta)^2$ and $\beta_c=1$ are two different bifurcation points. The action $\mathcal{S}$ shows different scaling laws with the distance of the infectious rate to the transition points $\beta_b$ and $\beta_c$, characterized by two different exponents: 3/2 and 1, respectively. Interestingly, the second-order derivative of $\mathcal{S}$ with respect to $\beta$ is discontinuous at $\beta=\beta_c$, while $\mathcal{S}$ and its first-order derivative are both continuous, reminiscent of the second-order phase transitions in equilibrium systems. Finally, a rare-event simulation method is used to compute the mean extinction time, which depends exponentially on $\mathcal{S}$ and the size $N$ of the population. The simulations results are in well agreement with the proposed theory. 
\end{abstract}
\maketitle

\section{Introduction}
In modern society, epidemic spreading, such as SARS, Ebola virus, COVID-19 has posed a serious threat to human health and global economic development. The extinction of epidemics is one of the major challenges in population dynamics \cite{Anderson1991,Nature460.334}. Although many factors may contribute, such as environmental changes and social factors, intrinsic fluctuations, originated from discreteness of the reacting agents and the random character of their interactions, can induce a rare but large fluctuation along a most probable, or optimal, path to extinction of epidemics \cite{PhysRevLett.101.078101,PhysRevLett.103.068101,PhysRevLett.101.268103,PhysRevE.81.021116}. Unlike equilibrium systems determined by the Boltzmann distribution, population dynamics is far away from equilibrium, and therefore there is no general principle to determine the probability of large fluctuations in out of equilibrium systems.

In the theoretical aspect, some mathematical epidemic models have been employed to study the problem of epidemic extinction. Such models, like the celebrated susceptible-infected-susceptible (SIS) model \cite{RevModPhys.87.925}, possess a stable density of infectious population when the infection rate exceeds an epidemic threshold. However, for a finite size population, the epidemic state is always metastable since a rare large fluctuation 
can bring it to an absorbing state where no infective individuals survive. Of two particular concerns are optimal path to extinction and mean time to extinction (MTE) (see \cite{WKBReview1} for a review and references therein). The starting point of theoretical studies is an exact master equation that describes the evolution of the underlying stochastic process. However, for most of systems the master equation cannot be exactly solved. When the size of population is large, the Fokker–Planck approximation to the master equation, via the van Kampen system size expansion or related recipes, only describes small deviations from the probability distribution maxima, but it fails to determine the probability of large fluctuations \cite{MultiscaleModelSimul3.283}. Elgart and Kamenev made an important progress in this topic,
pioneered in
Refs\cite{JSP9.51,PhysRevA.36.5782,Bull.Math.Biol.51.625,JCP100.5735}:
they employed the Peliti-Doi technique
\cite{JPA9.1465,JPhysFrance46.1469,PhysRevLett.77.4780,JSP90.1}, to
map the master equation into a Schr\"odinger-like equation that can
identify the classical trajectory that connects the metastable fixed
point and the absorbing state \cite{PhysRevE.70.041106}. This allows
them to calculate the classical action along this trajectory, a
first approximation to the MET. Assaf and Meerson then suggested a
general spectral method to improve the Elgart-Kamenev results
\cite{PhysRevLett.97.200602,PhysRevE.75.031122}. Kessler and Shnerb
presented a general method to deal with the extinction problem based
on the time-independent ``real space" Wentzel-Kramers-Brillouin
(WKB) approximation \cite{JSP127.861}. This method is easy to
implement, its intuitive meaning is transparent, and its range of
applicability covers any single species problem. These novel methods
have been successfully applied to solve extinction problems in
diverse situations, including time-varying environment
\cite{PhysRevE.78.041123,PhysRevE.81.031126,PhysRevLett.125.048105,PhysRevE.101.022109}, catastrophic events
\cite{PhysRevE.79.011127}, fragmented populations with migration
\cite{PhysRevLett.109.138104,PhysRevLett.109.248102,PhysRevE.101.012135}, complex
networks \cite{PhysRevE.88.012809,EPL108.58008,PhysRevLett.117.028302,PhysRevE.95.052317,PhysRevE.97.012308,PhysRevLett.123.068301}, and some others
\cite{PhysRevE.77.061107,schwartz2009predicting,PhysRevE.81.051925,PhysRevE.83.011129,PhysRevE.85.021140,PhysRevE.87.032127,PhysRevE.93.032109,chen2017epidemic,PhysRevE.99.022101,PhysRevLett.128.078301}.

Recently, a simplicial SIS model has been received increasing attention \cite{battiston2021physics}. The model can well describe the social contagion phenomena such as the adoption of norms, behaviors or new products, or the diffusion of rumors or fads. In contrast to the previous SIS model, the simplicial model does not only consider the pairwise transmission between a susceptible individual and an infectious one, but also incorporates higher-order interactions in a group of three or more individuals \cite{barrat2022social}. When a susceptible individual is exposed to multiple sources, the transmission can be reinforced by simplicial interactions associated with group
pressure. The model has been studied in simplicial complexes \cite{iacopini2019simplicial}, showing a
discontinuous transition from a healthy to endemic phase
when the relative weight of higher-order interactions crosses a threshold. Similar phenomena have also been observed
in heterogeneous \cite{PhysRevResearch.2.012049} and time-varying structures \cite{chowdhary2021simplicial,st2021universal}, and in the more general setup of hypergraphs \cite{jhun2019simplicial,de2020social,landry2020effect,st2022influential,PhysRevResearch.3.033282}.

However, the previous studies did not consider the problem of epidemic extinction in the simplicial SIS model. The epidemic extinction is key to understanding and controlling the dynamics of contagion in social systems, as mentioned before. In this work, we address this problem in a simplicial SIS model containing two-body and three body interactions. As a preliminary step, we do not consider the contact structure among individuals and thus assume that the population is well-mixed. As reported in previous studies, when the relative infection rate $\beta$ lies in between $\beta_b$ and $\beta_c$ the model show a bistable region where the endemic phase and healthy phase are coexisting. For $\beta>\beta_c$, only the endemic phase is stable. Due to the intrinsic stochasticity in a finite-size system, a rare fluctuation can bring the system from the endemic phase to the healthy phase, where the latter is an absorbing state implying that the system cannot leave once it enters into the state. For a large but finite system, there exist an optimal or the most probable path of epidemic extinction. Thanks to the  WKB approximation for the master equation, the problem is convert to find the zero-energy trajectories in an effective Hamiltonian system. The action along the optimal path gives the mean time to extinction. We find that the system possesses two radically different paths for $\beta_b<\beta<\beta_c$ and $\beta>\beta_c$. The action is also obtained explicitly. Interestingly, we find that the action and its first-order derivative with respect to $\beta$ are both continuous, but its second-order derivative is discontinuous at $\beta=\beta_c$, which is similar to continuous phase transitions in equilibrium systems where the second-order derivative of free-energy with respect to a control parameter is discontinuous when phase transition occurs. Furthermore, we find that the action shows the different scaling relations with the distance of the infectious rate to the transition points $\beta_b$ and $\beta_c$, characterized by two different exponents: 3/2 and 1, respectively.        

The rest of the paper is structured as follows. In Sec.\ref{sec2}, we define a simplicial SIS model and present a mean-field analysis for the model. In Sec.\ref{sec4}, we present the master equation for the stochastic description of the model and solve the master equation by WKB approximation. In Sec.\ref{sec5}, the optimal pathway of extinction and the associated action are given. Numerical validation for the theoretical results is presented in Sec.\ref{sec6}. Finally, main conclusions is provided in Sec.\ref{sec7}.

\section{Simplicial SIS model and Mean-field theory}\label{sec2}
Consider a well-mixed population consisted of $N$ individuals, in which each individual is either susceptible ($S$) or infected ($I$). 
The interactions between them can be described by the following reactions,
\begin{eqnarray}
S + I\xrightarrow{\beta }&I + I  \\
S + 2I\xrightarrow{{\beta ( {1 + \delta } )}} & I + 2I  \\
I\xrightarrow{\mu } &S   
\end{eqnarray}
where the first reaction accounts for an $S$ individual is infected with an infection rate $\beta$ when contacting with an infectious one. The second reaction corresponds to a contagion process within a group of three individuals, where one of individuals susceptible and the others are infectious. Under the case, the infection rate is $\beta (1+\delta)$ with $\delta>0$, which is larger than one-to-one contagion process due to the synergistic effect. The last reaction refers to the recovery process controlled by a recovery rate $\mu$. Without loss of generality, we set $\mu=1$ such that the infection rate is scaled with the recovery rate.  

\begin{figure}
	\centerline{\includegraphics*[width=1.0\columnwidth]{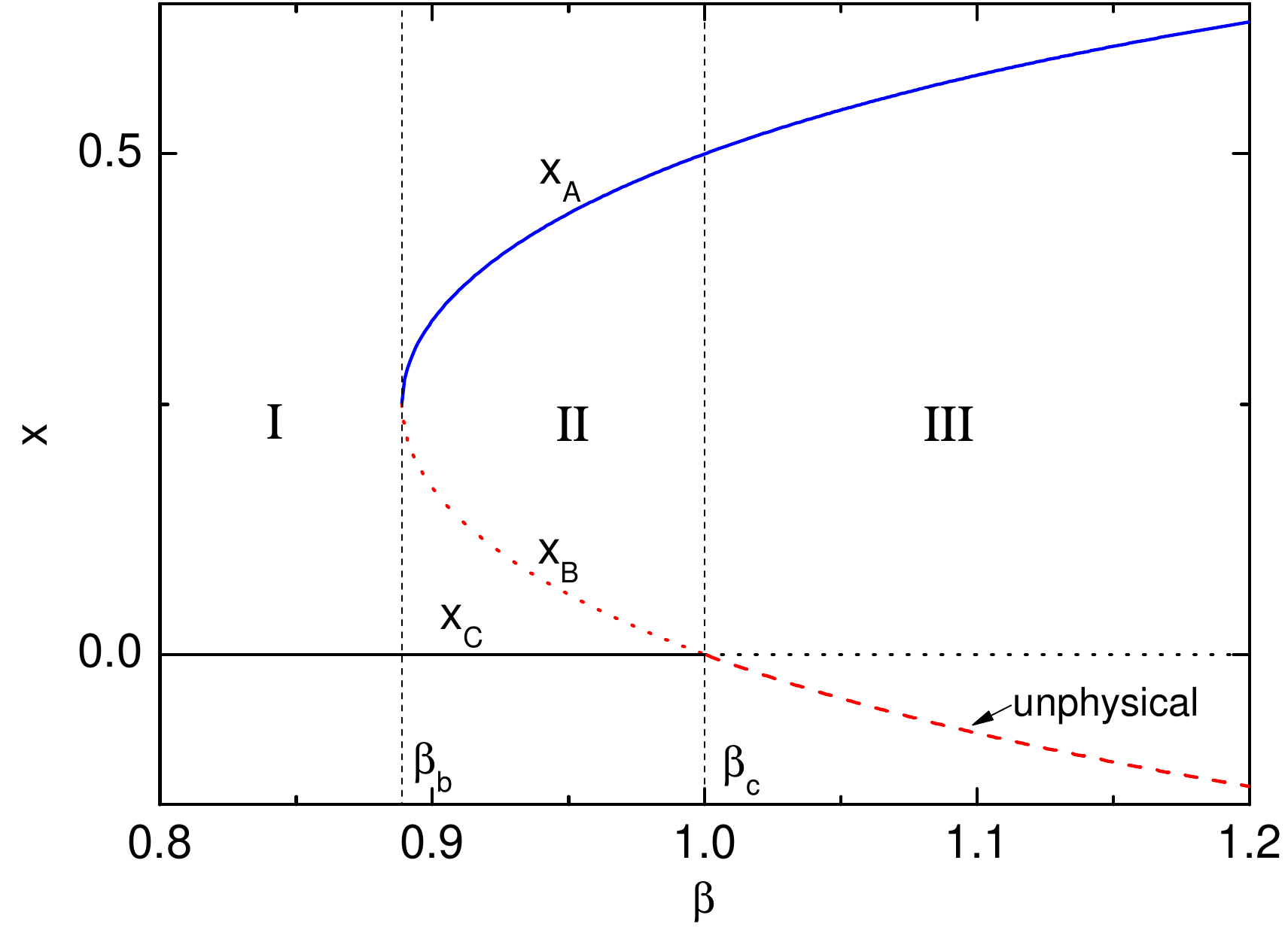}}
	\caption{ A bifurcation diagram of the simplicial SIS model for $\delta=1$. The bifurcation diagram is divided into three regions: (I) For $\beta<\beta_b=4 (1+\delta)/(2+\delta)^2$, there is no epidemic and the system is in the absorbing phase; (II) For $\beta_b<\beta<\beta_c=1$, the system is bistable in which the active phase and  absorbing phase are coexisting; (III) For $\beta>\beta_c$, only the active phase is stable.  \label{fig1}}
\end{figure}

Let us denote by $n$ the number of infectious individuals, and $x=n/N$ be the fraction of infectious individuals among the population. The rate equation for $x$ can be written as,
\begin{eqnarray}\label{eq15}
\dot x  = \beta x\left( {1 - x} \right)\left[ {1 + ( {1 + \delta } )x} \right] - x
\end{eqnarray}
Eq.(\ref{eq15}) has three fixed points, 
\begin{eqnarray}
{x_A} &=& \frac{\delta  + \sqrt{\Delta}}{{2\left( {1 + \delta } \right)}}\label{eq6}  \\
{x_B} &=& \frac{\delta  - \sqrt{\Delta}}{{2\left( {1 + \delta } \right)}}  \\
{x_C} &=& 0  
\end{eqnarray}
with $\Delta=( 2 + \delta  )^2 - 4( {1 + \delta } )/\beta$. However, $x_A$ and $x_B$ exist only for $\Delta \geq 0$, or equivalently, $\beta \geq \beta_b=4 (1+\delta)/(2+\delta)^2$, and they collapse at $\beta=\beta_b$ via a saddle-node bifurcation. According to linear stability analysis, $x_C$ is always stable as long as $\beta<\beta_c=1$. A bifurcation diagram for $\delta=1$ is plotted in Fig.\ref{fig1}.  The bifurcation diagram is divided into three regions.
\begin{itemize}
	\item[(I)] For $\beta<\beta_b$, both $x_A$ and $x_B$ do not exist, and $x_C$ is the only stable solution. Therefore, the number of infectious individuals decays exponentially and eventually the system enters into the absorbing phase where no infectious individual survive.
	\item[(II)] For $\beta_b<\beta<\beta_c$, both $x_A$ and $x_C$ are stable, separated by an unstable solution $x_B$. The system is bistable where the active phase and the absorbing phase are coexisting.    
	\item[(III)] For $\beta>\beta_c$, $x_A$ is stable, $x_C$ is unstable, while $x_B<0$ is an unphysical solution. The system is in the active phase.  
\end{itemize} 

We should note that the occurrence of Region (III) is due to the simultaneously incorporated effect of the two-body interactions and three-body interactions, which was not observed in a previous model where only the three-body interactions is present \cite{chen2017epidemic}. The difference leads to more abundant behaviors in the present model, which will be reported in the following.

\section{Master equation and WKB approximation}\label{sec4}
Mean-field theory predicts the behavior of an infinite system where the stochastic fluctuations can be ignored. However, for a finite size system the stochastic fluctuations is profound. This is because that the fluctuations can bring the population into the absorbing phase sooner or later where the epidemic is in extinction, in the sense that epidemic phase is always metastable. In the paper, our main goal is to calculate the mean time from the active phase to the absorbing phase. Such a mean extinction time is key as it measures the lifetime of the metastable epidemic phase.

To capture the effect of the stochastic fluctuations, we first define $P_n(t)$ as the probability that the system has $n$ infectious individuals at time $t$. The time evolution of $P_n(t)$ is governed by the master equation, 
\begin{eqnarray}\label{eq8}
\frac{{\partial {P_n}( t )}}{{\partial t}} &=& {W_ + }( {n - 1} ){P_{n - 1}}( t ) + {W_ - }( {n + 1} ){P_{n + 1}}( t ) \nonumber \\&-& \left[ {{W_ + }( n ) + {W_ - }( n )} \right]{P_n}( t )
\end{eqnarray}  
where $W_+(n)$ ($W_-(n)$) denotes the rate of the number of infectious individuals increased (decreased) by one providing that there is $n$ infectious individuals at present, given by \cite{van1992stochastic}
\begin{eqnarray}
{W_ + }\left( n \right) &=& \frac{{\beta \left( {N - n} \right)n}}{N} + \frac{{\beta ( {1 + \delta })\left( {N - n} \right)n\left( {n - 1} \right)}}{{N\left( {N - 1} \right)}}  \\
{W_ - }\left( n \right) &=& n  
\end{eqnarray}

As customary, we assume $N$ is large and take the leading order in an $N^{-1}$ expansion. This is similar to WKB ansatz of quantum mechanics, where $N^{-1}$ plays the role of Planck's constant in Schr\"odinger’s equation. By using
\begin{eqnarray}\label{eq10}
P_n(t)=e^{-N \mathcal{S}(x=n/N)}
\end{eqnarray}
and taking the leading order in $N^{-1}$, $P_{n \pm 1} \approx P_n e^{\mp \partial \mathcal{S}/ \partial x }$ and $W_{\pm}(n \pm 1)\approx W_{\pm}(n)$, the master equation \ref{eq8} can be converted to the Hamilton–Jacobi equation,
\begin{eqnarray}
\frac{{\partial \mathcal{S}}}{{\partial t}} + \mathcal{H}\left( {x,p} \right) = 0
\end{eqnarray}
where $\mathcal{S}$ and $\mathcal{H}$ are called the action and Hamiltonian, respectively. As in classical mechanics, the Hamiltonian is a function of the coordinate $x$ and its conjugate momentum $p = \partial \mathcal{S}/ \partial x$,
\begin{eqnarray}\label{eq11}
\mathcal{H}( {x,p} ) = {w_ + }( x )\left( {{e^p} - 1} \right) + {w_-}(x)\left( {{e^{ - p}} - 1} \right)
\end{eqnarray}
where
\begin{eqnarray}
{w_ + }( x ) &=& {{{W_ + }( n )}}/{N} \nonumber \\ &=& \beta x( {1 - x} ) + \beta ( {1 + \delta } ){x^2}( {1 - x} )  \\
{w_ - }( x ) &=& {{{W_ - }( n )}}/{N} = x  
\end{eqnarray}
are infection rate and recovery rate per individual, respectively.

The canonical equations of motion can be written as,
\begin{eqnarray}\label{eq12}
\dot x = {\partial _p}\mathcal{H}( {x,p} ) =  \beta x\left( {1 - x} \right)\left[ {1 + ( {1 + \delta } )x} \right]{e^p} - x{e^{ - p}} \nonumber \\
\end{eqnarray}
\begin{eqnarray}\label{eq13}
\dot p =  - {\partial _x}\mathcal{H}( {x,p} ) &=&  - \beta \left[ {1 - 2x + ( {1 + \delta } )( {2x - 3{x^2}} )} \right] \nonumber \\ &\times&  ( {{e^p} - 1} ) - {e^{ - p}} + 1 
\end{eqnarray}

We are interested in the extinction trajectory from an epidemic state to an extinct
state of epidemics. This means that there will be some trajectory along which $\mathcal{S}$ is minimized, which represents the most probable path of such an extinction event. This corresponds to the zero-energy ($\mathcal{H}=0$) trajectory in the phase space $(x,p)$ from an epidemic fixed point $A$ to an extinction one $C$. In terms of Eq.(\ref{eq11}), $\mathcal{H}=0$ requires that there are three lines: (i) extinction trajectory $x = 0$; (ii) mean-field trajectory $p = 0$, and (iii) activation trajectory 
\begin{eqnarray}\label{eq14}
p=p_a(x) = \ln \frac{{{w_-}(x)}}{{{w_ + }(x)}} =  - \ln \left[ {\beta \left( {1 - x} \right)\left( {1 + ( {1 + \delta } )x} \right)} \right] \nonumber \\
\end{eqnarray}
Such three trajectories determine the topology of the optimal extinction path on the phase plane $\left(x, p \right)$. In particular, the trajectory $p = 0$ corresponds to the result of mean-field treatment, as equation (\ref{eq12}) for $p = 0$ recovers to the mean-field equation (\ref{eq15}). \textcolor{blue}{That is to say, in the zero momentum subspace without fluctuations, it is impossible to bring the system escaped from an active phase to an extinction phase. The presence of nonzero momentums renders the escape event possible. In the subsequent section, we will show such an optimal escape path is controlled by a nonzero-momentum heteroclinic trajectory in the phase space. }

\begin{figure*}
	\centerline{\includegraphics*[width=1.8\columnwidth]{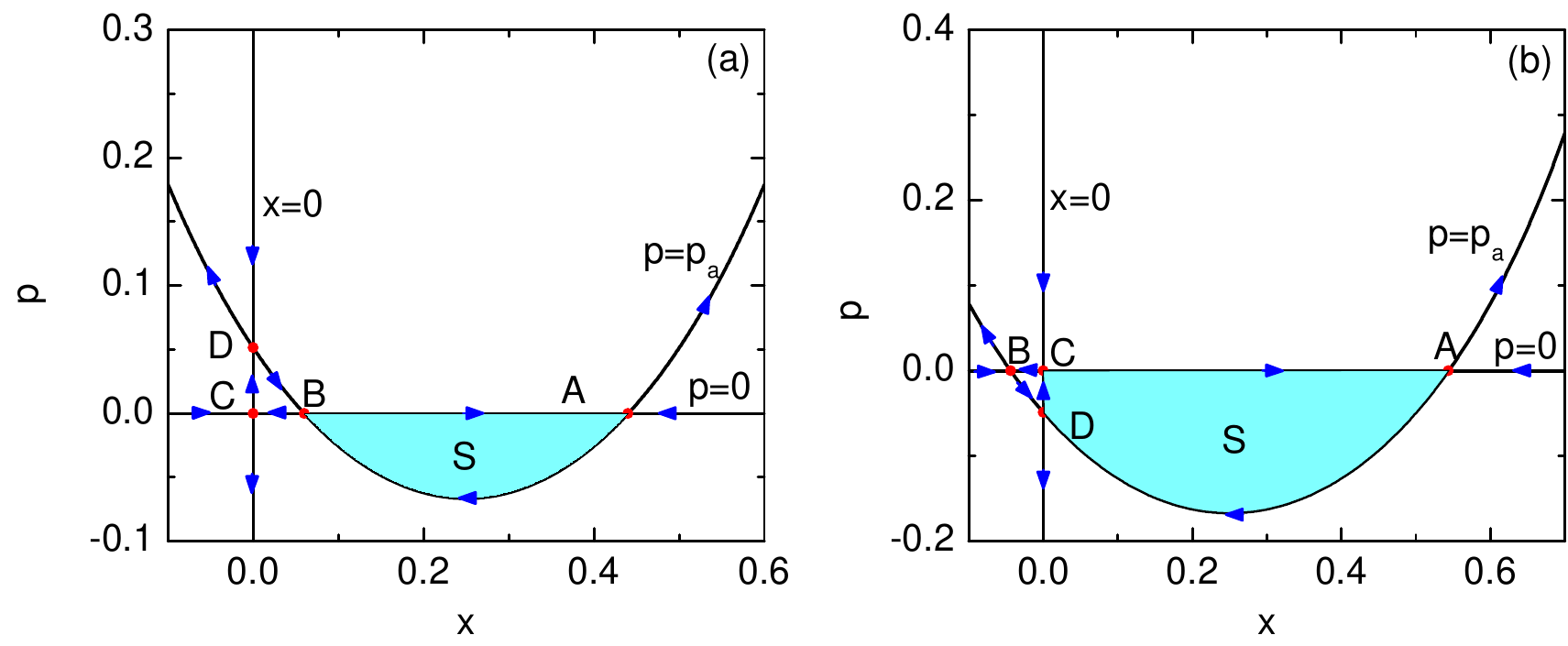}}
	\caption{Optimal path of extinction in region (II) (a) and in region (III) (b). The colored areas represent the action along the zero-energy trajectory from the fixed point $A$ to the fixed point $C$ in the $\left( x, p\right)$ phase space.  \label{fig2}}
\end{figure*}

\section{Optimal path to Epidemic Extinction}\label{sec5}
In the region (II) of Fig.\ref{fig1}, mean-field theory tells us that $x_C=0$ is an attracting fixed point of the rate equation (\ref{eq15}), so that the metastable population possesses an average fraction $x_A$ of susceptible individuals. In the stochastic description, extinction occurs via a large fluctuation which brings the population from $x_A$ to the repelling fixed point $x_B$. From there the system flows into the absorbing state $x_C= 0$ almost deterministically. In the framework of WKB theory the transition from $x_A$ to $x_C$ occurs in the extended phase plane $\left( x, p\right) $ where all three fixed points are hyperbolic, see Fig.\ref{fig2}(a). Here the optimal path to extinction is composed of two segments: the nonzero-momentum heteroclinic trajectory connecting the hyperbolic
fixed points $A=\left(x_A, 0 \right) $ and $B=\left(x_B, 0 \right) $ along the activation trajectory $p_a(x)$ in Eq.(\ref{eq14}), and the zero-momentum segment going from $B=\left(x_B, 0 \right) $ to $C=\left(x_C, 0 \right)$ via the relaxation trajectory. The action $\mathcal{S}$ along zero-energy trajectory is given by
\begin{eqnarray}\label{eq19}
\mathcal{S} (\beta,\delta) = F( {x_B} ) - F( {x_A} )
\end{eqnarray}
where
\begin{eqnarray}\label{eq20}
F( x ) &=& \int {{p_a}(x)dx }  \nonumber \\ =&&  2x + \ln ( {1 - x} ) - x\ln \left[ {\beta ( {1 - x} )( {1 + x + \delta x} )} \right] \nonumber \\ &-& \frac{1}{{1 + \delta }}\ln \left[ {1 + ( {1 + \delta } )x} \right]
\end{eqnarray}

For $\beta$ slightly larger than $\beta_b$, the action $\mathcal{S}$ scales with the distance of the infectious rate to the bifurcation point $\beta_b$. This can be obtained by a series expansion for $\mathcal{S}$ to the leading order in $\beta-\beta_b$, which yields 
\begin{eqnarray}\label{eq21}
\mathcal{S} (\beta,\delta) \sim \kappa_1 {\left( {\beta  - {\beta _b}} \right)^{\frac{3}{2}}}
\end{eqnarray}
where the prefactor $\kappa_1$ is dependent on the parameter $\delta$, given by 
\begin{eqnarray}
\kappa_1 = \frac{{{{( {2 + \delta } )}^4}}}{{12{{( {1 + \delta } )}^{5/2}}}}
\end{eqnarray}
The scaling exponent $3/2$ was also found in Ref.\cite{chen2017epidemic}, which seems to be universal near a saddle–node bifurcation point \cite{JCP100.5735}.  The exponent $3/2$ can be understood as follows. For $\beta=\beta_b$, $x_A=x_B$ and the two fixed points $A$ and $B$ shown in Fig.\ref{fig2}(a) collide with each other. For $\beta \gtrsim \beta_b$, the heteroclinic path from $A$ to $B$ can be regarded as two straight lines connected by a lowest point, where coordinates of the lowest point in the $\left(x,p \right) $ phase space can be located by the activation trajectory in Eq.(\ref{eq14}), given by $\left(\frac{\delta}{2(1+\delta)},  - \ln \left( {\beta /{\beta _b}} \right) \right) $. Therefore, the action is approximately computed as $\mathcal{S}\approx \frac{1}{2}(x_A-x_B) \ln \left( {\beta /{\beta _b}} \right)=\frac{{\sqrt {\beta _b^{ - 1} - {\beta ^{ - 1}}} }}{{\sqrt {\left( {1 + \delta } \right)} }} \ln \left( {\beta /{\beta _b}} \right) \sim  {\left( {\beta  - {\beta _b}} \right)^{\frac{3}{2}}}$.

In the region (III), $x_C$ becomes a repelling fixed point of the rate equation (\ref{eq15}). In a stochastic description extinction occurs via a large fluctuation which, acting against an effective entropy barrier, brings the population from $x_A$ directly to the absorbing state $x_C=0$. In the WKB language this transition is possible  because of the presence of the fluctuational extinction point, $D$, in an extend phase space $\left(x ,p \right) $, see Fig.(\ref{fig2}(b)). \textcolor{blue}{Here, the fluctuational extinction point $D$ is the intersection point of the extinction trajectory ($x=0$) and activation trajectory ($p=p_a(x)$, see also Eq.(\ref{eq14})), from which one can easily determine the coordinates of $D$, $D=\left(0, -\ln \beta \right) $. }
The most probable path to extinction is the heteroclinic trajectory connecting the metastable point $A=\left( x_A, 0 \right) $ and the fluctuational extinction point $D$, and then to the absorbing point $C=\left(0, 0 \right) $. The action $\mathcal{S}$ along the zero-energy trajectory is given by
\begin{eqnarray}\label{eq23}
\mathcal{S} (\beta,\delta) = F( {x_C} ) - F( {x_A} )
\end{eqnarray}
where $F(\cdot)$ is also given in Eq.(\ref{eq20}).

For $\beta$ slightly larger than $\beta_c$, we perform a series expansion for $\mathcal{S}$, and find that the difference $\mathcal{S}(\beta,\delta)-\mathcal{S}(\beta_c,\delta)$ scales with the distance of the infectious rate to $\beta_c$, given by
\begin{eqnarray}\label{eq24}
\mathcal{S}(\beta,\delta)-\mathcal{S}(\beta_c,\delta) \sim \kappa_2 \left( {\beta-{\beta _c}} \right)
\end{eqnarray}
with the prefactor  
\begin{eqnarray}
	\kappa_2 = \frac{\delta}{1+\delta}
\end{eqnarray}

To the best of our knowledge, the scaling exponent $1$ has not yet been reported in the previous literature. The exponent is in contrast to the case for the standard scaling of the activation of escape near a transcritical bifurcation, which is $2$ for the latter \cite{PhysRevLett.101.078101,PhysRevE.77.061107,schwartz2009predicting}. The exponent $1$ stems from the topology of the optimal path to extinction depicted in Fig.\ref{fig2}(b). For $\beta=\beta_c$, the fluctuational extinction point $D=\left(0, -\ln \beta \right) $ collides with the absorbing point $C=\left(0, 0\right) $. As $\beta$ increases from  $\beta_c$, the point $D$ moves down along the $p$-axis and the point $A=\left(x_A, 0 \right) $ moves right along the $x$-axis. For $\beta \gtrsim \beta_c$, the amounts of these two movememts are both $\beta-\beta_c$, and thus the additional area enclosed by the optimal extinction path is given by Eq.(\ref{eq24}).

\begin{figure}
		\centerline{\includegraphics*[width=1.0\columnwidth]{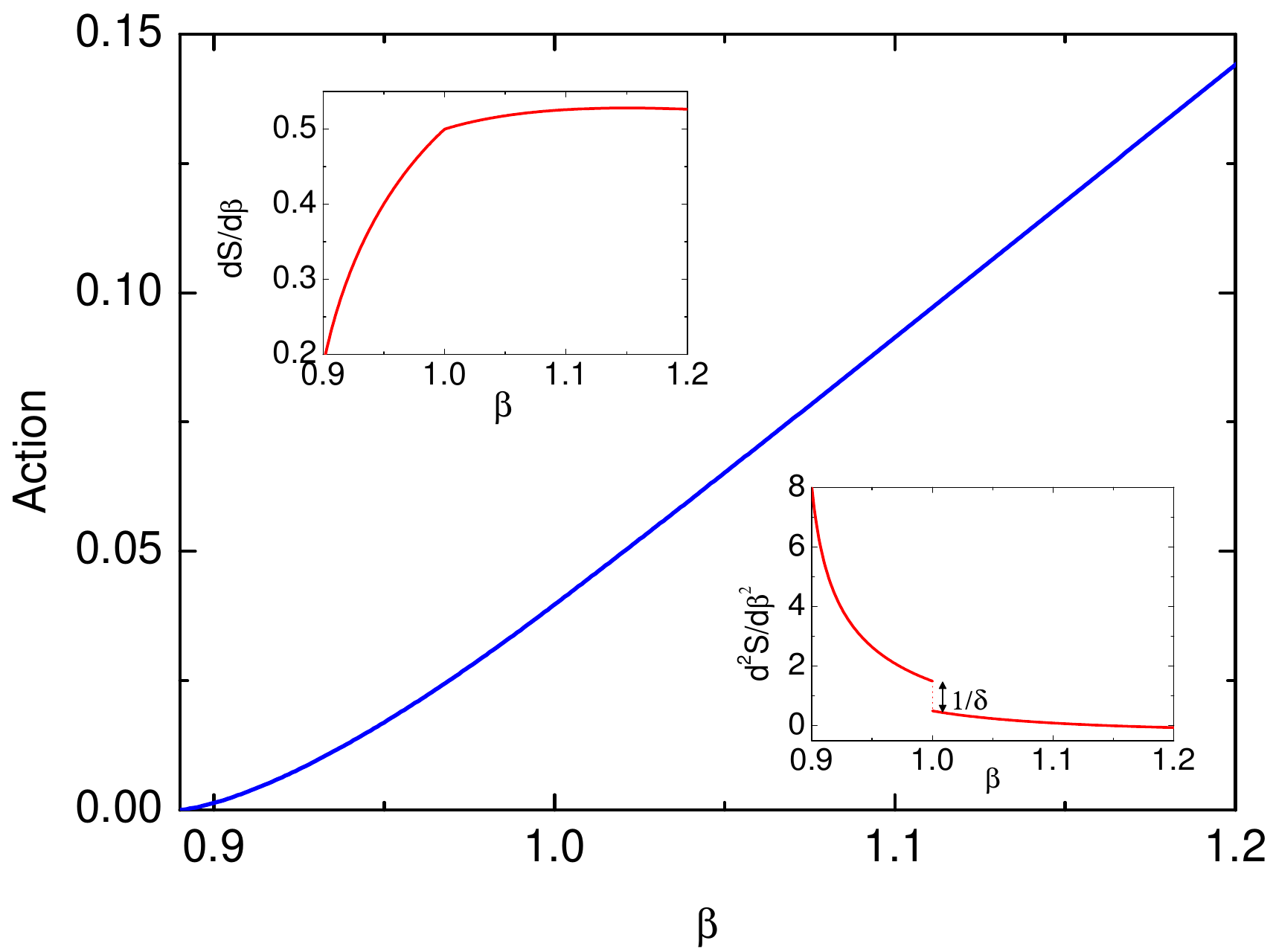}}
		\caption{ Action as a function of $\beta$ for $\delta=1$. The insets show the first-order and the second-order derivatives of the action with respect to $\beta$, where the second-order derivative is discontinuous at $\beta=\beta_c=1$.   \label{fig3}}
\end{figure}

It is useful to summarize the main results in the present work. For $\beta_c<\beta<\beta_b$ and $\beta>\beta_c$, we have revealed the two different scenarios for optimal path to extinction, and obtained the action $\mathcal{S}$ along the optimal path for each scenario. In Fig.\ref{fig3}, we show $\mathcal{S}$ as a function of $\beta$ for $\beta>\beta_b$, where $\delta$ is fixed at $\delta=1$. In the insets of Fig.\ref{fig3}, we also show the first-order and the second-order derivatives of $\mathcal{S}$ with respect to $\beta$ as a function of $\beta$. Interestingly, we find that $\mathcal{S}$ and its first-order derivative with respect to $\beta$ are both continuous at $\beta=\beta_c$, but the second-order derivative is discontinuous at $\beta=\beta_c$. After cumbersome calculations, the discontinuity is given by
\begin{eqnarray}\label{eq26}
	\left.  \frac{\partial \mathcal{S}}{\partial \beta}\right| _{\beta=\beta_c^-}- \left. \frac{\partial \mathcal{S}}{\partial \beta} \right| _{\beta=\beta_c^+}   =\frac{1}{\delta}
\end{eqnarray}
	
The discontinuity of the second-order derivative of $\mathcal{S}$ with respect to $\beta$ at $\beta=\beta_c$ is reminiscent of the second-order phase transitions in equilibrium systems. This is because that the action plays a similar role of the free-energy in equilibrium systems, which measures the mean lifetime of the active phase or the relative stability of the phase. From Eq.(\ref{eq26}), one can see that the degree of the discontinuity at $\beta=\beta_c$ decreases as $\delta$ increases. In the limit of $\delta \to \infty$, the model is totally dominated by the three-body interactions and the discontinuity vanishes. Thus one can conclude that the discontinuity in the second-order derivative of $\mathcal{S}$ is due to the combination of the two-body interactions and three-body interactions. \textcolor{blue}{The occurrence of the singularity in the action is often called \textit{dynamical} phase transition \cite{touchette2009large}. The mechanism of the dynamical phase transition in the present work is purely originated from the geometric change of the optimal extinction path at $\beta=\beta_c$, as shown in Fig.\ref{fig2}. The transition appears only when the size of population is large enough such that the WKB ansatz is justified, and is finite such that the fluctuation is present to drive the extinction event.}

From Eq.(\ref{eq10}), the extinction rate is determined by the action calculated for $x\to 0$, i.e., by the probability density for reaching the disease-free state. 
It is easy to see that the minimum of the action is realized by the optimal extinction path, as shown Fig.\ref{fig2}(a) and Fig.\ref{fig2}(b) for $\beta_b<\beta<\beta_c$ and for $\beta>\beta_c$, respectively. Thus the entropic barrier for extinction is $N \mathcal{S}$, and the mean extinction time is given by \cite{PhysRevLett.101.078101}
\begin{equation}
\langle T \rangle  = e^{N \mathcal{S}}
\end{equation}
where $\mathcal{S}$ is computed in terms of Eq.(\ref{eq19}) and Eq.(\ref{eq23}) for $\beta_b<\beta<\beta_c$ and for $\beta>\beta_c$, respectively.
 
\section{Numerical validation}\label{sec6}
In order to validate the theoretical results, we have performed the
stochastic simulation for the master equation (\ref{eq8}) by Gillespie's
algorithm \cite{JCP22.403,JPC81.2340}. However, epidemic extinction is a rare event that
occurs very infrequently, especially for large $\beta$ or $N$.
Thus, the conventional brute-force simulation becomes prohibitively
inefficient. To overcome this difficulty, we have employed an efficient rare-event sampling method, forward flux sampling (FFS)
\cite{PRL05018104,JPH09463102}, combined with Gillespie's
algorithm. The FFS uses a series of interfaces between the initial and final states to calculate rate constants (or mean transition time) and generate transition paths, for rare events in equilibrium or nonequilibrium systems with stochastic dynamics. In one of our previous papers, we have used the FFS to obtain the mean time to extinction in a generalized SIS model, and the details of the method can be found there \cite{chen2017epidemic}.

\begin{figure}
	\centerline{\includegraphics*[width=1.0\columnwidth]{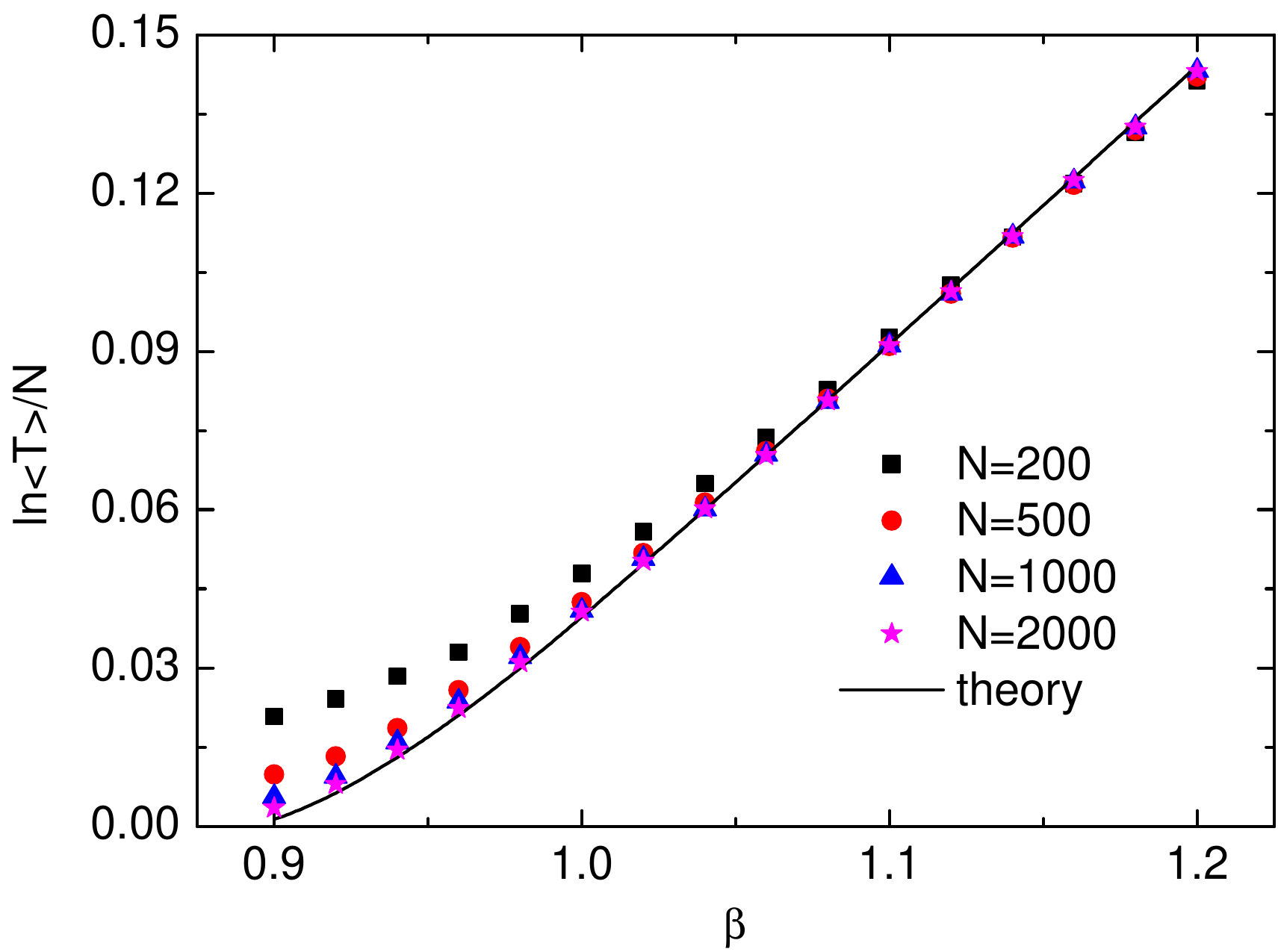}}
	\caption{ $\ln \langle T \rangle /N$ as a function of $\beta$ for $\delta=1$. Symbols correspond to simulation results and line to theory. \label{fig4}}
\end{figure}

In Fig.\ref{fig4}, we show $\ln \langle T \rangle /N$ as a function of $\beta$ for different size $N$ of population with a fixed $\delta=1$. For large $N$, there is excellent agreement between the simulations
and theoretical predictions for all $\beta$'s. For small $N$, the agreements hold only for large values of $\beta$. 
While for small $N$ and $\beta$, the theory disagrees the simulations. This is because that in this case $ \langle T \rangle $ is not very long, such that  $e^{N \mathcal{S}}$ can be comparable to its pre-exponential factor which was not considered in our analysis. 

\begin{figure}
	\centerline{\includegraphics*[width=1.0\columnwidth]{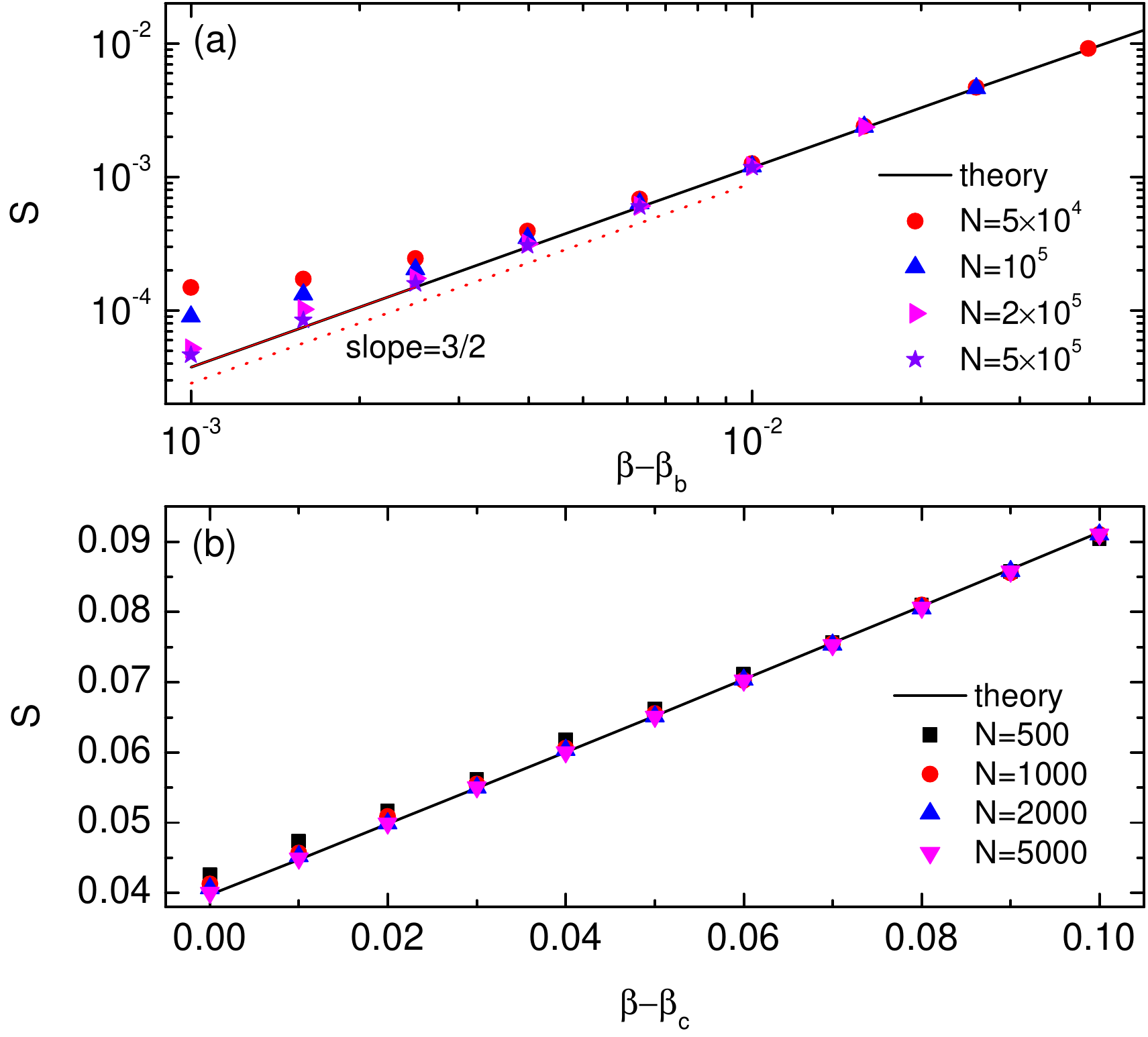}}
	\caption{ The action as a function of the distance of $\beta$ from the bifurcation points $\beta_b$ (a) and $\beta_c$ (b). The other parameter is $\delta=1$. Symbols correspond to simulation results and line to theory. \label{fig5}}
\end{figure}

Finally, we shall show the numerical verification of the scaling relation
of the action near two bifurcation points, $\beta_b$ and $\beta_c$, as shown in Eq.(\ref{eq21}) and Eq.(\ref{eq24}), respectively.  In Fig.\ref{fig5}, we compare the simulation and theoretical results for different $N$'s. Clearly, the simulation results support our theoretical predictions.

\section{Conclusions}\label{sec7}
In conclusion, we have studied the epidemic extinction in a simplicial SIS model, where a high-order interaction between susceptible and infectious individuals is involved. By employing WKB approximation for the master equation, the study is converted to finding the zero-energy trajectory in a Hamilton system, and the associated action along the optimal extinction path is related to the mean time to extinction. Depending on the scaled infection rate $\beta$, we have revealed two different most likely paths from an epidemic phase to the epidemic extinction phase for $\beta_b<\beta<\beta_c$ and for $\beta>\beta_c$, and the corresponding action is also obtained explicitly. Interestingly, we find that the action and its first-order derivative with respect to $\beta$ are both continuous, but the second-order derivative of action with respect to $\beta$ is discontinuous at $\beta=\beta_c$, which is reminiscent of the second-order phase transitions in equilibrium systems. Furthermore, we show that the action near $\beta_b$ has a scaling form with an exponent 3/2, and the difference between the action and the action at $\beta=\beta_c$ has also a scaling relation with a different exponent 1. All the theoretical results are verified by the stochastic simulations combined with a rare-event sampling method. Our study unveils an interesting mechanism of a rare or large fluctuation on a nonequilibrium system when high-order interactions are considered. In the future, it would be desirable
to consider the epidemic extinction of the simplicial SIS model on complex interacting topologies, which can be conveniently represented by simplicial complexes or hypergraphs.

\begin{acknowledgments}
This work is supported by the National Natural Science
Foundation of China (Grants Nos. 11875069, 11975025 and 12011530158).
C. S. was also funded by the Key Laboratory of Modeling,
Simulation and Control of Complex Ecosystem in Dabie
Mountains of Anhui Higher Education Institutes, and the
International Joint Research Center of Simulation and Control
for Population Ecology of Yangtze River in Anhui.
\end{acknowledgments}

\end{document}